\documentclass[12pt]{JHEP3}

\let\ifpdf\relax
\usepackage{ifpdf}
\usepackage{graphics}
\usepackage{graphicx}
\usepackage{amssymb}
\usepackage{amsmath}
\usepackage{slashed}
\usepackage{cite}
\usepackage{epsfig}
\usepackage{datetime}
\usepackage{epic}

\newcommand{\be}{\begin{equation}}
\newcommand{\ee}{\end{equation}}
\newcommand{\bal}{\begin{align}}
\newcommand{\eal}{\end{align}}
\newcommand{\bea}{\begin{eqnarray}}
\newcommand{\eea}{\end{eqnarray}}
\newcommand{\half}{\frac{1}{2}}

\def\a{\alpha}
\def\b{\beta}

\def\g{\gamma}

\def\e{{\epsilon}}

\def\half{\frac{1}{2}}

\def\Tr{{\rm Tr}}

\def\eps{{\epsilon}}

\usepackage{epic}
\def\option{\noindent}
\def\vs{\vspace}
\def\Tr{{\rm Tr}}

\def\e{{\rm e}}
\def\ie{{\it i.e.}}

\newcommand{\Z}{\mathsf{Z}\kern -5pt \mathsf{Z}}
\def\sig{\sigma}

\def\ishimu{  | \mu \rangle\!\rangle_I }

\def\id{ 0 }
\def\Pplus{ P_+^K }
\def\Smat{ S_{\mu\nu} }
\def\tSmat{ \tS_{\tmu\tnu} }
\def\Nabc{ {N_{\mu\nu}}^{\lam}  }

\def\tNabc{ {{\tN}_{\tmu \tnu}}^{~~\sig^\Dr(\tlam) } }

\def\om{\omega}
\def\omc{{\omega_c}}

\def\Pplus{ P_+^K }
\def\zeroth{zero${}^{\rm th}$ }

\def\pirho{{\pi(\rho)}}
\def\half{ {1\over 2} }

\def\id{ 0 }
\def\a{\alpha}
\def\b{\beta}
\def\Dr{{ \Delta }}

\def\lam{\lambda}
\def\sig{\sigma}
\def\om{\omega}
\def\Zcl{ Z^{\rm closed} }
\def\Zop{ Z^{\rm open} }

\def\Pplus{ P_+^K }
\def\zeroth{zero${}^{\rm th}$ }

\def\cH{  {\cal H}  }
\def\Exp{  {\cal E}^{\om}  }

\def\Bound{  {\cal B}^{\om}  }
\def\Boundomc{  {\cal B}^{\omc}  }
\def\tq{{\tilde{q}}}
\def\trho{{\tilde{\rho}}}
\def\ta{{\tilde{\a}}}

\def\tb{{\tilde{\b}}}
\def\tn{{\tilde{n}}}
\def\tN{{\tilde{N}}}

\def\tS{{\tilde{S}}}

\def\tlam{{\tilde{\lambda}}}

\def\tmu{{\tilde{\mu}}}
\def\tnu{{\tilde{\nu}}}
\def\tpsi{{\tilde{\psi}}}

\def\rhohat{{\hat{\rho}}}
\def\pirhohat{{\pi(\rhohat)}}
\def\hS{{S'}}
\def\htS{{\tS'}}
\def\htSstar{{\tS^{\prime *}}}

\def\bz{{\bar z}}

\def\bV{\overline{V}}
\def\bJ{\overline{J}}
\def\bL{\overline{L}}

\def\bJ{\overline{J}}
\def\g{g}
\def\gcup{ {\breve g} }
\def\sun{{{\rm su}(N)}}

\def\sukn{{\widehat{\rm su}(K)_N}}
\def\suN{{{\rm su}(N)}}
\def\sun{{{\rm su}(n+1)}}

\def\spn{{{\rm sp}(n)}}

\def\spNK{{\widehat{\rm sp}(N)_K}}
\def\spKN{{\widehat{\rm sp}(K)_N}}
\def\hgK{{\hat{g}_K}}
\def\g{g}
\def\gcup{ {\breve g} }
\def\suN{{{\rm su}(N)}}
\def\sun{{{\rm su}(n+1)}}

\def\spn{{{\rm sp}(n)}}

\def\hgK{{\hat{g}_K}}
\def\twhgK{{\hat{g}_K^\om}}
\def\twhgk{{\hat{g}_{2k+1}^\omc}}
\def\sunk{{\widehat{\rm su}(N)_K}}
\def\suNK{{\widehat{\rm su}(N)_K}}
\def\sonk{{\widehat{\rm so}(N)_K}}

\def\sunK{{\widehat{\rm su}(n+1)_K}}

\def\suoddnk{{\widehat{\rm su}(2n+1)_{2k+1}}}
\def\suoddkn{{\widehat{\rm su}(2k+1)_{2n+1}}}

\def\suKN{{\widehat{\rm su}(K)_N}}

\def\AnoneK{  (A_{n}^{(1)})_K}
\def\Annonek{  (A_{2n}^{(1)})_{2k+1}}
\def\Anntwok{  (A_{2n}^{(2)})_{2k+1}}
\def\spnK{{\widehat{\rm sp}(n)_K}}
\def\spnk{{\widehat{\rm sp}(n)_k}}
\def\spkn{{\widehat{\rm sp}(k)_n}}
\def\ConeK{  (C_n^{(1)})_K}
\def\Conek{  (C_n^{(1)})_k}
\def\twhgK{{\hat{g}_K^\om}}
\def\twhgk{{\hat{g}_{2k+1}^\omc}}
\def\suNK{{\widehat{\rm su}(N)_K}}
\def\sunK{{\widehat{\rm su}(n+1)_K}}
\def\suKN{{\widehat{\rm su}(K)_N}}

\def\spnK{{\widehat{\rm sp}(n)_K}}
\def\spnk{{\widehat{\rm sp}(n)_k}}
\def\spkn{{\widehat{\rm sp}(k)_n}}

\def\AnoneK{  (A_{n}^{(1)})_K}

\def\ConeK{  (C_n^{(1)})_K}
\def\Conek{  (C_n^{(1)})_k}

\def\bdy{  | B \rangle\!\rangle }
\def\twbdy{  | B \rangle\!\rangle^\om }

\def\ishimu{  | \mu \rangle\!\rangle_I }
\def\twishimu{  | \mu \rangle\!\rangle_I^\om }
\def\Ctwishimu{  | \mu \rangle\!\rangle_I^\omc }
\def\ishinu{  | \nu \rangle\!\rangle_I }
\def\twishinu{  | \nu \rangle\!\rangle_I^\om }
\def\ishimubra{  {}_I\langle \! \langle  \mu | }
\def\twishimubra{  {}_I^\om\langle \! \langle  \mu | }
\def\cardymu{  | \mu \rangle\!\rangle_C }
\def\cardya{  | \a \rangle\!\rangle_C^\om }
\def\Ccardya{  | \a \rangle\!\rangle_C^\omc }
\def\cardyb{  | \b \rangle\!\rangle_C^\om }
\def\cardylam{  | \lam \rangle\!\rangle_C }
\def\cardylambra{  {}_C \langle \! \langle  \lam  | }
\def\cardyabra{  {}_C^\om \langle \! \langle  \a  | }

\def\nblama{ {n_{\b\lam}}^{\a}  }

\def\Nabc{ {N_{\mu\nu}}^{\lam}  }

\def\tNabcsig{ {{\tN}_{\tmu \tnu}}^{~~\tlam } }

\def\nabc{ {n_{\mu\nu}}^{\lam}  }
\def\four{  {\vcenter  {\vbox
			{\hrule height.4pt
				\hbox {\vrule width.4pt  height3pt
					\kern3pt
					\vrule width.4pt  height3pt
					\kern3pt
					\vrule width.4pt  height3pt
					\kern3pt
					\vrule width.4pt  height3pt
					\kern3pt
					\vrule width.4pt height3pt}
				\hrule height.4pt}
		}
	}
}
\def\oneoneoneone{
	{\vcenter  {\vbox
			{\hrule height.4pt
				\hbox {\vrule width.4pt  height3pt
					\kern3pt
					\vrule width.4pt  height3pt }
				\hrule height.4pt
				\hbox {\vrule width.4pt  height3pt
					\kern3pt
					\vrule width.4pt  height3pt }
				\hrule height.4pt
				\hbox {\vrule width.4pt  height3pt
					\kern3pt
					\vrule width.4pt  height3pt }
				\hrule height.4pt
				\hbox {\vrule width.4pt  height3pt
					\kern3pt
					\vrule width.4pt  height3pt }
				\hrule height.4pt}
		}
	}
}
\def\two{  {\vcenter  {\vbox
			{\hrule height.4pt
				\hbox {\vrule width.4pt  height3pt
					\kern3pt
					\vrule width.4pt  height3pt
					\kern3pt
					\vrule width.4pt height3pt}
				\hrule height.4pt}
		}
	}
}
\def\oneone{
	{\vcenter  {\vbox
			{\hrule height.4pt
				\hbox {\vrule width.4pt  height3pt
					\kern3pt
					\vrule width.4pt  height3pt }
				\hrule height.4pt
				\hbox {\vrule width.4pt  height3pt
					\kern3pt
					\vrule width.4pt  height3pt }
				\hrule height.4pt}
		}
	}
}

\title{Left-Right Entanglement Entropy, D-Branes, and Level-rank duality}

\author{ Howard J. Schnitzer  \\
{\small  Martin Fisher School of Physics, Brandeis University, \\ \ \ \ \ \ Waltham, MA 02454, USA\\
}

E-mail:
 \email{schnitzr@brandeis.edu}}

\preprint{
 BRX-TH-6296}

\abstract{We consider the left-right entanglement (LREE) entropy in 1+1 dimensions for WZW models on a circle, and for WZW models on untwisted and twisted D-branes. The consequences of level-rank duality for these applications  is presented which provides a map of LREE from large to small central charge.}

\begin{document}
	
	\section{Introduction}
	
	Entanglement entropy  has become an important topic of interest in quantum field theory, AdS/CFT correspondence, black-hole physics, and many other applications. For systems where the Hilbert space can be written as $\mathcal{H} = \mathcal{H_{A}}\bigotimes\mathcal{H_{B}}$ , the quantum entanglement of the subsystem $\mathcal{H_{A}}$  is obtained from the reduced density matrix $\rho_{A}$, and the von Neumann entropy $S_{A} = -tr(\rho_{A} ln\rho_{A})$, computed by tracing over the degrees of freedom of $\mathcal{H_{B}}$.

An interesting application of this strategy is that of a class of excited states of CFT on a circle in $1+1$ dimensions with conformally invariant boundary states $|B \rangle \rangle$ which are various linear combinations of Ishibashi states $|\lambda\rangle\rangle_{I}$\cite{Ishibashi:1988kg} appropriate for a CFT on a circle in $1+1$ dimensions,\cite{PandoZayas:2014wsa,Das:2015oha} D-branes in string theories, and D-branes for WZW models. These are examples which have a basis for the factorization into left and right moving sectors. One can trace over the right-moving states to obtain a left-right entanglement entropy(LREE).

Recently the formalism to calculate the LREE for a CFT on a circle in $1+1$ dimensions was developed\cite{Das:2015oha}. However Ishibashi states do not satisfy the appropriate boundary conditions in many cases. For a CFT on a finite cylinder, WZW models on D-branes, and other applications, Cardy states are appropriate\cite{Cardy:1989ir}. These are
\bea
|\lambda \rangle \rangle_C=\sum_{\mu}\frac{S_{\lambda \mu}}{\sqrt{S_{0\mu}}}|\lambda \rangle \rangle_I
\eea
where $S_{\lambda\mu}$ is the modular transformation matrix of the CFT. For example, the LREE for Cardy states on a circle in $1+1$ is\cite{Das:2015oha}

\bea
S_{|\lam  \rangle\rangle_{C}} =\frac{\pi c l}{24 \epsilon } -\sum_{\mu}S^2_{\lambda \mu}\ln \left(\frac{S^2_{\lambda \mu}}{S_{0\mu}}\right)
\eea
where  $\textit{l}$ is  the circumference of the circle, and $\epsilon$ is an UV cutoff introduced to define regularized boundary states.
The Cardy states play a prominent role in this paper as will become evident.

We will be concerned with the WZW models for the classical Lie algebras, $\sunk$ and $\spNK$. These WZW models have important algebraic properties, level-rank duality\footnote{It is a common misconception that level-rank duality involves a map from a Young tableau to another one obtained by interchanging rows and columns of the tableau. Rather, it is a one to one  map of equivalence classes. See appendices},\cite{Naculich:1990hg, Altschuler:1989nm, Mlawer:1990uv, Naculich:2005tn, Naculich:2006mt} which
involve maps of cominimal  classes of
\bea
\sunk\rightarrow \sukn \quad \textrm{and} \quad \spNK \rightarrow  \spKN .
\eea
Spinors representations of $\sonk$ involve several complications for level-rank duality of $\sonk$, so that $\sonk$ will not be discussed in this paper.

We will present consequences of level-rank duality for LREE for two applications; $1)$ WZW models on the circle in $1+1$ dimensions. and $2)$ untwisted and twisted D-branes of WZW models.\cite{Naculich:2005tn, Naculich:2006mt} The latter are concerned with strings propagating on a group manifold.

We give an extensive review of level-rank duality as applied to WZW models on untwisted and twisted D-branes in the Appendices. These provide the necessary information needed for the body of the paper.

It was shown\cite{Das:2015oha} that when the boundary states on the circle in $1+1$ dimensions are chosen to be Ishibashi states, the LREE entropy reproduces the spatial entanglement entropy of a $(2+1)$ dimensional topological field theory (TQFT).\footnote{See figure 1 of \cite{Das:2015oha}.} For a Cardy state on a circle, the LREE can be interpreted as the average topological entanglement entropy. Extending this, it would be interesting if there was a topological interpretation in $(2+1)$ dimensions corresponding to the WZW models on untwisted and twisted D-branes. One possibility is that of Wilson line observables of Chern-Simons theories on $S^{3}$, with the gauge group of an associated WZW model.

Recall that the central charge for $\sunk$ is
\bea
\label{ccsunk}
c = \frac{K (N^2-1)}{K+N}
\eea

and for $\spnk$
\bea
\label{ccsupnk}
c = \frac{K N(N+1)}{(N+K+1)}
\eea
Consider the limit $K\rightarrow \infty$, N fixed for (\ref{ccsunk}) and (\ref{ccsupnk}), so  that
\bea
c \sim (N^2-1) \quad \textrm{and}\quad c\sim N(N+1) \quad \textrm{respectively.}
\eea
Then for the level-rank duals $\sukn$ and $\spKN$
\bea
c \sim K N,
\eea
again for N fixed, $K\rightarrow \infty$.

If there is a topological description for the examples of this paper, perturbatively it could be described by an expansion in $1/c$, in the large $c$ limit, which would be a strong-coupling limit of the CFT. Then by using the level rank duality, one could evaluate the WZW models in a weak-coupling expansion in $(1+1)$ dimensions, and compare with the topological theory in detail.

\section{LREE on a circle and level-rank duality}

In this section we consider the consequences of level-rank duality for the LREE of WZW models on the $1+1$ dimensional circle of circumference $l$. The necessary formalism was developed in \cite{PandoZayas:2014wsa,Das:2015oha}, so that here we only select results needed for our discussion. Consider an arbitrary boundary state in terms of Ishibashi states
\bea
|B \rangle \rangle =\sum_{\lambda} \psi_B^{\lambda} |h_\lam \rangle \rangle_I
\eea
where $\lam$ is the label for the $\lam^{th}$ primary of the CFT with weight $h_\lam$. The state $|B \rangle \rangle $ is non-normalizable, so that a regularized state\cite{PandoZayas:2014wsa,Das:2015oha}
\bea
|B \rangle \rangle = \frac{e^{-\eps H}}{\sqrt{N_B}} |B \rangle \rangle
\eea
is introduced, with $N_B$ a normalization factor. The Hamiltonian is
\bea
H=\frac{2\pi}{l}(L_0+\bar{L}_0-\frac{c}{12})
\eea
for a circle of circumference $l$, $L_0$ a Virasoro generator, c the central charge of the CFT, and $\eps$ a cut-off to be considered for $\eps\to 0$.

In \cite{Das:2015oha} it is shown that the LREE is\footnote{We use the conventional notation $S_{0 \lam}$ to denote the modular transformation matrix connected to the identity}

\bea
\label{SBLREE}
S_{|B  \rangle\rangle} =\frac{\pi c l}{24 \epsilon } -\frac{\sum_{\lam}S^2_{0 \lambda } |\psi_B^{\lam}|^2 \ln \left(|\psi_B^{\lam}|^2\right)}{\sum_{\lam}S^2_{0 \lambda } |\psi_B^{\lam}|^2}+ \ln \left(\sum_{\lam} S_{0\lam }|\psi_B^{\lam}|^2\right)
\eea
where $S_{\mu \lambda }$ is the modular transformation matrix of the CFT. For our applications, we concentrate on Cardy states. [see Appendices (\ref{A1}) to (\ref{A4}) for details]. Untwisted Cardy states \footnote{Twisted Cardy states are discussed in Appendix \ref{A3} in the context of twisted D -branes } can be written as
\bea
|\lambda \rangle \rangle_C=\sum_{\mu}\frac{S_{\lambda \mu}}{\sqrt{S_{0\mu}}}|\mu \rangle \rangle_I
\eea
for generic Cardy states in a diagonal CFT, the LREE is
\bea
\label{sl}
S_{|\lam  \rangle\rangle} =\frac{\pi c l}{24 \epsilon } -\sum_{\mu}|S_{\lambda \mu}|^2\ln \left(\frac{|S_{\lambda \mu}|^2}{|S_{0\mu}|}\right)
\eea

Now specialize (\ref{sl}) to $\sunk$ and $\spNK $ WZW models. The level-rank dual of the divergent part of (\ref{sl}) is trivially given by  (\ref{ccsunk}) and (\ref{ccsupnk}). Therefore we concentrate on the finite part of (\ref{sl}) in what follows.

Consider the untwisted Cardy state $|0 \rangle \rangle_{C}$ for the identity representation. Then the LREE is\cite{Das:2015oha}
\bea
\label{lree}
S_{|0 \rangle \rangle_{C}} = \frac{\pi c l}{24 \epsilon} -\sum_{\lambda} (\frac{\textbf{d}_{\lambda}}{D})^2 \ln ((\frac{\textbf{d}_{\lambda}}{D})
\eea
where $\textbf{d}_{\lambda} = (\frac{S_{0 \lambda}}{S_{00}})$ is the quantum dimension of the conformal primary $\lambda$, and
\bea
D = [ \sum_{\lambda} \textbf{d}_{\lambda}^{2} ]^{1/2} = \frac{1}{S_{00}}
\eea
is the total quantum dimension. It is shown in \cite{Naculich:1990hg}  that
\bea
\label{dual}
(\frac{S_{0 \lambda}}{S_{00}})_{G(N)_{K}} = (\frac{\tilde{S}_{0 \tilde{\lambda}}}{\tilde{S}_{00}})_{G(K)_{N}}
\eea
for all \underline{non-spinor} representations of all classical groups, for all N and K. Since the sum in (\ref{lree}) for $\hat{so}(N)_{K}$ includes spinor representations, level-rank duality is considerable more complicated, so that $\hat{so}(N)_{K}$  will not be considered in this paper. Combining (\ref{lree}) to (\ref{dual}) and (\ref{eq:sss}) we conclude that $(S_{|0 \rangle \rangle_{C}})_{finite}$ is invariant for
$\spNK \leftrightarrow  \spKN$.  For $\sunk\leftrightarrow \sukn$, we  turn to generic Cardy states $|\lambda \rangle \rangle_{C}$.

\section*{$\sunk\leftrightarrow \sukn$ } 

From (\ref{eq:smunu}), the finite part of (\ref{sl}) for $\sunk$, expressed in terms of the S-matrices of $\sukn$ is
\bea
- \sum_{\mu \epsilon P^{K}_{+}} ( \frac{K}{N}) |\tilde{S}_{\tilde{\lambda }\tilde{\mu}}|^{2} \ln \left \{ {\sqrt{\frac{K}{N}}\frac{|\tilde{S}_{\tilde{\lambda }\tilde{\mu}}|^{2}}{|\tilde{S}_{0 \tilde{\mu}}|}} \right \}
\eea
where $\tilde{S}_{\tilde{\lambda}\tilde{\mu}}$ is the modular transformation matrix of $\sunk$. The sum $\mu$ over the primary states of $\sunk$ is equivalent to N\,times the sum over cominimal equivalence classes of $\sunk$. These in turn are in 1 to 1 correspondence with the cominimal equivalence classes of $\sukn$ and finally a factor of 1/K to sum over primary states of $\sukn$. Therefore we conclude that
 \bea
 \left( S_{{ | \lambda\rangle\rangle}_{c}} \right)_{finite} = -\frac{1}{2} \ln \left( \frac{K}{N}\right) + \left( \tilde{S}_{{ | \tilde{\lambda}\rangle\rangle}_{c}} \right)_{finite}
\eea
which relates the (finite part) of LREE of $\sunk$ to that of $\sukn$.

\section*{ $\spnk\leftrightarrow \spkn$}

A similar calculation using (\ref{eq:sss}) shows that for this case
\bea
 \left( S_{{ | \lambda\rangle\rangle}_{c}} \right)_{finite} =   \left( \tilde{S}_{{ | \tilde{\lambda}\rangle\rangle}_{c}} \right)_{finite}
\eea
\section{LREE for untwisted D-branes of WZW models}
The basic description of untwisted D-branes for WZW models is reviewed in Appendix B, where the closed-string Hamiltonian
\bea
H = \half \left( L_0 + \bL_0 - \frac 1{12} c \right)
\eea
appears in (\ref{eq:ishinorm})ff. A LREE is obtained from a cylinder with boundary conditions $\alpha$ and $\beta$ on the two boundaries. One constructs
a reduced density matrix $\rho_{A}$ by tracing over the degrees of freedom on boundary $\beta$. As explained in Appendix B,
the coherent states $\|B\rangle\rangle$ for this set-up are given by Cardy states $\|\lambda\rangle\rangle_{c}$ c.f. equation
 (\ref{eq:cardyishi}). The derivation of \cite{Das:2015oha} is applicable with the change $l \rightarrow  4 \pi$ in (\ref{sl}). Thus
\bea
S_{|\lam  \rangle\rangle} =\frac{\pi^2 c }{6 \epsilon } -\sum_{\mu\epsilon P^{K}_{+}}|S_{\lambda \mu}|^2\ln
 \left(\frac{|S_{\lambda \mu}|^2}{|S_{0\mu}|}\right)
\eea
Therefore the consequences of level-rank duality is identical to that of section II.

\section{LREE for twisted D-branes   of  ${\widehat{\rm su}(2n+1)_{2k+1}}$ WZW models}
The $\omega$-twisted Cardy states, described in Appendix C, are
\bea
\label{lwc}
|\lambda\rangle\rangle_{c}^{\omega} = \sum_{\mu\,\epsilon\,\varepsilon^{\omega}}
\frac{\psi_{\lambda\pi(\mu)}}{\sqrt{S_{0\mu}}}|\mu\rangle\rangle_{I}^{\omega}
\eea
where $\psi_{\lambda\pi(\mu)}$ is identified with the modular transformation matrix of $(A_{2n}^{(2)})_{2k+1}$, and $S_{0\lambda}$ that of a self-conjugate
 representation of
${\widehat{\rm su}(2n+1)_{2k+1}}$. Further, as discussed in Appendix C, one can identify
\bea
\label{psialpha}
\psi_{\lambda\pi(\mu)} = S'_{\lambda\pi(\mu)}
\eea
as the modular transformation matrix of $\spNK$.
From (\ref{lwc}) and (\ref{psialpha}), together  with (\ref{SBLREE}) we obtain the LREE for the  $\omega$-twisted D-branes
\bea
\label{slwc}
S_{|\lambda\rangle\rangle_{c}^{\omega}} = \frac{\pi^2 c}{4\epsilon} - \sum_{\mu\,\epsilon\,\varepsilon^{\omega}}
|S'_{\lambda\pi(\mu)}|^2 \ln[\frac{S'_{\lambda\pi(\mu)}|^2}{|S_{0\mu}|}]
\eea
where the modular properties of $S'_{\lambda\pi(\mu)}\,\,\epsilon\,\,\spNK$ have been used.

We now turn to the level-rank duality of the finite part of (\ref{slwc}). From
(\ref{eq:squrt})
\bea
S'_{\alpha\beta} =  \tilde{S}'_{\tilde{\alpha}'\tilde{\beta}'} = \tilde{S}'^{*}_{\tilde{\alpha}'\tilde{\beta}'}
\eea
where $S'$ and $\tilde{S}'$ denote the modular transform matrices of $\spNK$ and $\spKN$ respectively, which involves a one-to-one map
of the primary states of
$\spNK$ to those of  $\spKN$. Since $S_{0\mu}$ is that of a self-conjugate representation of ${\widehat{\rm su}(2n+1)_{2k+1}}$,
it is related to that of
${\widehat{\rm su}(2k+1)_{2n+1}}$ by (\ref{eq:smunu}). Therefore
\bea
|S_{0\mu}| = (\frac{2 k +1}{2 n +1})^{\frac{1}{2}} |\tilde{S}_{0\tilde{\mu}}|
\eea
Since
\bea
\sum_{\mu\,\epsilon\,\varepsilon^{\omega}}|S'_{\lambda\pi(\mu)}|^2 = 1
\eea
we obtain the level-rank dual of the twisted LREE
\bea
[S_{|\lambda\rangle\rangle_{c}^{\omega}}]_{finite} = [\tilde{S}_{|\tilde{\lambda}\rangle\rangle_{c}^{\omega}}]_{finite} 
+\dfrac{1}{2} \ln(\frac{2k+1}{2n+1})
\eea
where the sum $\mu\, \varepsilon\, \epsilon^{\omega}$ is one-to-one with the sum $\tilde{\mu}\,\epsilon\,\varepsilon^{\omega}$ as is appropriate to
$\spNK\rightarrow\spKN$

\section{Concluding Remarks}
The LREE of WZW models on the 1 + 1 dimensional circle and on untwisted and  twisted D-branes was presented, together with the consequences of level-rank duality. The limit $K \rightarrow \infty$, N fixed, relates dual theories, one with large central charge to one with finite central charge. This can be useful in considering strong coupling limits. It is known that the LREE CFT on a circle can be related to that of a topological theory in 2 + 1 dimensions. It is an open question if there is a topological interpretation in 2 + 1 dimensions for the LREE for WZW models on untwisted and twisted branes.  Considerations of these issues is in progress.

\section*{Acknowledgements}
We wish to thank Steve Naculich for his collaboration on a long series of papers on which the results of this paper depend and for reading the manuscript. We also thank C. Ag\'on and I. Cohen for their invaluable help in preparing this paper and M. Headrick for his comments.
H.J.S. is supported in part by the DOE by grant DE-SC0009987.
	
\appendix
\section{Level-rank duality overview \label{A1}}

\setcounter{equation}{0}
\label{secintro}

Level-rank duality is a relationship between
various quantities in bulk Wess-Zumino-Witten models
with classical Lie groups \cite{
	Naculich:1990hg, 
	Altschuler:1989nm, 
	Mlawer:1990uv}.
It  has  been shown \cite{
	Naculich:2005tn,
	Naculich:2006mt}
that level-rank duality also applies to untwisted
and to certain twisted D-branes
in the corresponding boundary WZW models
(For a review of D-branes on group manifolds  see \cite{Klimcik:1996hp,Behrend:1998fd,Felder:1999ka,Stanciu:1999id,Fredenhagen:2000ei,Maldacena:2001ky,Gawedzki:2001ye,Ishikawa:2001zu,Bouwknegt:2002bq,Schomerus:2002dc,Freed:2001jd,Ishibashi:1988kg,Fuchs:1995zr,Kacbuk,Goddard:1986bp,Fuchsbuch,Bouwknegt:2006pd,Alekseev:1998mc,Verlinde:1988sn,Bouwknegt:2000qt}
Untwisted (i.e., symmetry-preserving) D-branes of WZW models
are labelled by the integrable highest-weight
representations $V_\lam$ of the affine Lie algebra.
For example, for $\suNK$,
these representations belong to cominimal equivalence classes \cite{Naculich:1990hg, Mlawer:1990uv}
generated by the $\Z_N$ simple current of the WZW model,
and therefore so do the untwisted D-branes of the model.
Level-rank duality is a one to one correspondence
between cominimal equivalence classes (or simple-current orbits)
of integrable representations of $\suNK$ and $\suKN$,
and therefore induces a map between
cominimal equivalence classes
of untwisted D-branes.

The spectrum of an open string stretched between
D-branes labelled by $\a$ and $\b$
is specified by the coefficients of the partition function\cite{Naculich:2005tn, Naculich:2006mt}
\be
\label{eq:openpartition}
\Zop_{\a\b} (\tau) = \sum_{\lam \in \Pplus} \nblama \chi_\lam (\tau)
\ee
where $\chi_\lam (\tau)$
is the affine character of the integrable highest-weight
representation $V_\lam$.
For untwisted D-branes discussed in Appendix B, the coefficients $\nblama$ are equal to
the fusion coefficients of the bulk WZW theory \cite{Cardy:1989ir},
so the well-known level-rank duality of the fusion
rules \cite{
	Naculich:1990hg,
	Altschuler:1989nm,
	Mlawer:1990uv}
implies the duality of the open-string spectrum between untwisted branes. For twisted D-branes, see Appendices C and D.

\section{Untwisted D-branes of WZW models \label{A2}}
\renewcommand{\theequation}{B.\arabic{equation}}
\setcounter{equation}{0}

We review some salient features
of Wess-Zumino-Witten models
and their untwisted D-branes.\cite{Naculich:2005tn}

The WZW model,
which describes strings propagating on a group manifold,
is a rational conformal field theory
whose chiral algebra (for both left- and right-movers)
is the (untwisted) affine Lie algebra $\hgK$ at level $K$.
The Dynkin diagram of $\hgK$ has one more node
than that of the associated finite-dimensional Lie algebra $\g$.
Let $(m_0, m_1, \cdots, m_n)$ be the dual Coxeter labels of $\hgK$
(where $n = {\rm rank~} \g$)
and $h^\vee = \sum_{i=0}^n m_i$ the dual Coxeter number of $\g$.
The Virasoro central charge  of the WZW model is then
$c ={K \dim \g}/(K+h^\vee)$. For $\sunk$; $c = \frac{K (N^{2}-1)}{K + N}$, and for $\spNK$; $c = \frac{K N (N+1)}{(N+K+1)}$.

The building blocks of the WZW conformal field theory
are integrable highest-weight representations
$V_\lam$ of $\hgK$,
that is, representations whose highest weight $\lam \in \Pplus$
has non-negative Dynkin indices $(a_0, a_1, \cdots, a_n)$
satisfying
\be
\label{eq:integrable}
\sum_{i=0}^n  m_i a_i = K\,.
\ee
With a slight abuse of notation, we also use $\lam$ to denote
the highest weight of the irreducible representation of $\g$
with Dynkin indices $(a_1, \cdots, a_n)$,
which spans the lowest-conformal-weight subspace of $V_\lam$.

For $\sunK = \AnoneK$ and $\spnK = \ConeK$,
the untwisted affine Lie algebras with which
we will be concerned in this section, have
$m_i=1$ for $i=0, \cdots, n$, and $h^\vee = n+1$.
It is often useful to describe irreducible representations
of $\g$ in terms of Young tableaux.
For example,
an irreducible representation of $\sun$ or $\spn$
whose highest weight $\lam$ has Dynkin indices $a_i$
corresponds to a Young tableau with $n$ or fewer rows,
with row lengths
\be
\ell_i = \sum_{j=i}^{n} a_j \,, \qquad
i=1, \ldots, n  \,.
\ee
Let $r(\lam) = \sum_{i=1}^{n} \ell_i$ denote the number of boxes
of the tableau.
Representations $\lam$
corresponding to integrable highest-weight representations
$V_\lam$ of $\sunK$ or ~$\spnK$
have Young tableaux with $K$ or fewer columns.

We  consider in this section WZW theories with a diagonal closed-string spectrum:
\be
\label{eq:diagonal}
\cH^{\rm closed}
= \bigoplus_{\lam \in \Pplus}
V_\lam \otimes \bV_{\lam^*}
\ee
where $\bV$ represents right-moving states,
and $\lam^*$ denotes the representation conjugate to $\lam$.
The partition function for this theory is
\be
\label{eq:closedpartition}
\Zcl (\tau)
= \sum_{\lam \in \Pplus} \left|  \chi_\lam (\tau)  \right|^2
\ee
where
\be
\chi_\lam (\tau) = \Tr_{V_\lam} q^{L_0 - c/24}\,, \qquad
q =  \e^{2\pi i \tau}
\ee
is the affine character of the integrable highest-weight
representation $V_\lam$.
The affine characters transform linearly
under the modular transformation $\tau \to -1/\tau$,
\be
\label{eq:modulartrans}
\chi_\lam(-1/\tau) = \sum_{\mu \in \Pplus} S_{\mu\lam} \; \chi_\mu(\tau)\,,
\ee
and the unitarity of $S$ ensures
the modular invariance of the partition function
(\ref{eq:closedpartition}).

Next we consider D-branes in the WZW model\cite{Naculich:2005tn,
Klimcik:1996hp}.
These D-branes may be studied algebraically
in terms of the possible boundary conditions
that can consistently be imposed on a WZW model with boundary.
In this section we consider boundary conditions that leave unbroken the $\hgK$ symmetry,
as well as the conformal symmetry, of the theory,
and we label the allowed boundary conditions
(and therefore the D-branes) by $\a$, $\b$, $\cdots$.
The partition function on a cylinder,
with boundary conditions $\a$ and $\b$ on the two boundary components,
is then given as a linear combination of
affine characters of $\hgK$ \cite{Cardy:1989ir}
\be
\label{eq:bdypartition}
\Zop_{\a\b} (\tau) = \sum_{\lam \in \Pplus} \nblama \chi_\lam (\tau) \,.
\ee
This describes the spectrum of
an open string stretched between D-branes labelled by $\a$ and $\b$.

In this section,
we consider a special class of boundary conditions,
called {\it untwisted} (or {\it symmetry-preserving}),
that result from imposing the restriction
\be
\label{eq:untwistedconditions}
\left[ J^a(z) - \bJ^a(\bz)\right] \bigg|_{z=\bz} = 0
\ee
on the currents of the affine Lie algebra on the boundary $z=\bz$
of the open string world-sheet,
which has been conformally transformed to the upper half plane.
Open-closed string duality allows one to correlate
the boundary conditions (\ref{eq:untwistedconditions})
of the boundary WZW model
with coherent states $\bdy \in \cH^{\rm closed}$
of the bulk WZW model satisfying
\be
\label{eq:modes}
\left[  J^a_m + \bJ^a_{-m} \right] \bdy = 0\,, \qquad m\in \Z
\ee
where $J^a_m$ are the modes of the affine Lie algebra generators.
Solutions of eq.~(\ref{eq:modes})
that belong to a single sector $V_\mu \otimes \bV_{\mu^*}$
of the bulk WZW theory
are known as Ishibashi states $\ishimu$ \cite{Ishibashi:1988kg},
and are normalized such that
\be
\label{eq:ishinorm}
\ishimubra q^H \ishinu =  \delta_{\mu\nu} \chi_\mu (\tau)\,,
\qquad q = \e^{2\pi i \tau}
\ee
where $H = \half \left( L_0 + \bL_0 - \frac 1{12} c \right)$
is the closed-string Hamiltonian.
For the diagonal theory (\ref{eq:diagonal}),
Ishibashi states exist for all integrable highest-weight
representations $\mu \in \Pplus$ of $\hgK$.

A coherent state $\bdy$ that corresponds to an
allowed boundary condition
must also satisfy additional (Cardy) conditions \cite{Cardy:1989ir},
among which are that the coefficients $\nblama$ in
eq.~(\ref{eq:bdypartition}) must be non-negative integers.
Solutions to these conditions
are labelled by integrable highest-weight representations
$\lam \in \Pplus$ of the untwisted affine Lie algebra $\hgK$,
and are known as (untwisted) Cardy states $\cardylam$.
The Cardy states may be expressed as linear combinations
of Ishibashi states
\be
\label{eq:cardyishi}
\cardylam = \sum_{\mu \in \Pplus}
{S_{\lam \mu} \over \sqrt{S_{\id\mu}}} \ishimu
\ee
where $S_{\lam\mu}$ is the modular transformation matrix
given by eq.~(\ref{eq:modulartrans}),
and $\id$ denotes the identity representation.
Untwisted D-branes of $\hgK$ correspond to $\cardylam$
and are therefore also labelled by $\lam \in \Pplus$.

The partition function of open strings stretched between
untwisted D-branes $\lam$ and $\mu$
\be
\label{eq:openpart}
\Zop_{\lam\mu} (\tau) = \sum_{\nu \in \Pplus} \nabc \chi_\nu (\tau)
\ee
may alternatively be calculated as the closed-string propagator between
untwisted Cardy states \cite{Cardy:1989ir}
\be
\label{eq:prop}
\Zop_{\lam\mu} (\tau) = \cardylambra \tq^H \cardymu\,,
\qquad \tq = \e^{2\pi i (-1/\tau)   }\,.
\ee
Combining eqs.~(\ref{eq:prop}),
(\ref{eq:cardyishi}),
(\ref{eq:ishinorm}),
(\ref{eq:modulartrans}),
and the Verlinde formula \cite{Verlinde:1988sn},
we find
\be
\label{eq:verlinde}
\Zop_{\lam\mu} (\tau)
= \sum_{\rho \in \Pplus}
{ S^*_{\lam\rho} S_{\mu\rho} \over S_{\id\rho} } \chi_\rho(-1/\tau)
= \sum_{\nu \in \Pplus} \sum_{\rho \in \Pplus}
{ S_{\mu\rho} S_{\nu\rho} S^*_{\lam\rho} \over S_{\id\rho} }  \chi_\nu(\tau)
= \sum_{\nu \in \Pplus} \Nabc \chi_\nu(\tau)\,.
\ee
Hence, the coefficients $\nabc$ in the
open-string partition function  (\ref{eq:openpart})
are simply given by the fusion coefficients $\Nabc$ of the bulk WZW model.

\vs{.1in}
\noindent{\bf Level-rank duality of the untwisted open string spectrum of  $\sunk$}
\vs{.1in}

\option
A  relation, level-rank duality,
exists between the WZW model for $\sunk$
and the corresponding WZW model with $N$ and $K$
exchanged \cite{Naculich:1990hg, Altschuler:1989nm, Mlawer:1990uv}.
The Young tableau $\lam$ corresponding to an integrable highest-weight
representation of $\sunk$ maps under transposition
(\ie, exchange of rows and columns) to a Young tableau $\tlam$
that corresponds to an integrable highest-weight
representation of $\sukn$
(possibly after removing any columns of length $K$).
This map is not one-to-one,
since cominimally-equivalent representations of $\sunk$
may map into the same representation of $\sukn$
(due to the removal of columns).
It is clear, however, that the cominimal equivalence classes
of the two theories {\it are} in one-to-one correspondence.

The modular transformation matrices and fusion rule coefficients
of the $\sunk$ theory
obey simple relations under the exchange of $N$ and $K$.
Letting $\Smat$ and $\tSmat$ denote the modular transformation
matrices of $\sunk$ and $\sukn$, one finds \cite{Mlawer:1990uv}
\be
\label{eq:smunu}
\Smat = \sqrt {K\over N} \ \e^{-2\pi i  r(\mu) r(\nu)/NK} \ \tSmat^* ~.
\ee
where $r(\mu)$ is the number of boxes in the Young tableau associated to the representation $\mu$.
From this and eq.~(\ref{eq:verlinde}),
it follows that \cite{Mlawer:1990uv}
\be
\label{eq:fusionduality}
\Nabc  = \tNabc \,,
\qquad\qquad
\Dr =  {r(\mu) + r(\nu) - r(\lam) \over N} \in \Z
\ee
where $\tN$ denotes the fusion rule multiplicities of $\sukn$.
(For $N$ sufficiently large, i.e.,
 for $N > k_1 (\mu) + k_1 (\nu) $,
where $k_1(\mu)$ denotes the length of the first column  of $\mu$
$\Dr$ vanishes,
so that on the right-hand side of the fusion algebra (\ref{eq:verlinde}),
$\lam$ is simply dual to $\tlam$, its transpose,
but in general the relation is more complicated.)

\vs{.1in}
\noindent{\bf Level-rank duality of the untwisted open string spectrum of  $\spnk$}
\vs{.1in}

\option
In ref.~\cite{Mlawer:1990uv},
it was shown that the fusion coefficients $\Nabc$ of
the bulk $\spnk$ WZW model are related to those
of the $\spkn$ WZW model by
\be
\Nabc =  \tNabcsig\,.
\ee
Since the fusion coefficients $\Nabc$ are equal
to the coefficients $\nabc$ of the open-string partition function,
it follows that if
the spectrum of an $\spnk$ open string
stretched between untwisted D-branes $\lam$ and $\mu$
contains $\nabc$ copies of the
highest-weight representation $V_\nu$ of $\spnk$, then
the spectrum of an $\spkn$ open string
stretched between untwisted D-branes $\tlam$ and $\tmu$
contains an equal number of copies of the
highest-weight representation $V_{\tnu}$ of $\spkn$.

\section{Twisted D-branes of WZW models \label{A3}}
\renewcommand{\theequation}{C.\arabic{equation}}
\setcounter{equation}{0}

In this section we review some aspects of twisted D-branes of the WZW model,
drawing on \cite{Naculich:2006mt}.
As in section B,
these D-branes correspond to possible boundary conditions
that can be imposed on a boundary WZW model.

A boundary condition more general than eq.~(\ref{eq:untwistedconditions})
that still preserves the $\hgK$ symmetry of the boundary WZW model
is
\be
\label{eq:twconditions}
\left[ J^a(z) - \omega \bJ^a(\bz)\right] \bigg|_{z=\bz} = 0\,,
\ee
where $\omega$ is an automorphism of the Lie algebra $\g$.
The boundary conditions (\ref{eq:twconditions})
correspond to coherent states $\twbdy \in \cH^{\rm closed}$
of the bulk WZW model
that satisfy
\be
\label{eq:twmodes}
\left[  J^a_m + \omega \bJ^a_{-m} \right] \twbdy = 0\,, \qquad m\in \Z\,.
\ee
The $\om$-twisted Ishibashi states  $\twishimu$ are solutions
of eq.~(\ref{eq:twmodes})
that belong to a single sector
$ V_\mu \otimes \bV_{\om(\mu)^*} $
of the bulk WZW theory,
and whose normalization is given by
\be
\label{eq:twishinorm}
\twishimubra q^H \twishinu =  \delta_{\mu\nu} \chi_\mu (\tau)\,,
\qquad q = \e^{2\pi i \tau} \,.
\ee
Since we are considering the diagonal closed-string theory
(\ref{eq:diagonal}),
these states only exist when $\mu = \om(\mu)$,
so the $\om$-twisted Ishibashi states are labelled by
$\mu \in \Exp$, where $\Exp \subset \Pplus$ are the integrable
highest-weight representations of $\hgK$ that satisfy $\om(\mu)=\mu$.
Equivalently, $\mu$ corresponds to a highest-weight representation,
which we denote by $\pi(\mu)$,
of $\gcup$,
the orbit Lie algebra \cite{Kacbuk} associated with $\hgK$.

Solutions of eq.~(\ref{eq:twmodes}) that also satisfy the
Cardy conditions are denoted $\om$-twisted Cardy states $\cardya$,
where the labels $\a$ take values in some set $\Bound$.
The $\om$-twisted Cardy states may be expressed
as linear combinations of $\om$-twisted Ishibashi states
\be
\label{eq:twcardyishi}
\cardya = \sum_{\mu \in \Exp}
{\psi_{\a \pi(\mu)} \over \sqrt{S_{\id\mu}}} \twishimu
\ee
where $\psi_{\a \pi(\mu)}$ are some as-yet-undetermined coefficients.
The $\om$-twisted D-branes of $\hgK$ correspond to $\cardya$
and are therefore also labelled by $\a \in \Bound$.
These states  correspond
to integrable highest-weight representations
of the $\om$-twisted affine Lie algebra $\twhgK$.

The partition function
of open strings stretched between $\om$-twisted D-branes $\a$ and $\b$
\be
\label{eq:twopenpartition}
\Zop_{\a\b} (\tau) = \sum_{\lam \in \Pplus} \nblama \chi_\lam (\tau)
\ee
may alternatively be calculated as the closed-string propagator between
$\om$-twisted Cardy states
\be
\label{eq:twprop}
\Zop_{\a\b} (\tau) = \cardyabra \tq^H \cardyb\,,
\qquad \tq = \e^{2\pi i (-1/\tau)   }\,.
\ee
Combining eqs.~(\ref{eq:twprop}),
(\ref{eq:twcardyishi}),
(\ref{eq:twishinorm}),  and
(\ref{eq:modulartrans}),
we find
\be
\Zop_{\a\b} (\tau)
= \sum_{\rho \in \Exp}
{ \psi^*_{\a\pirho} \psi_{\b\pirho} \over S_{\id\rho} } \chi_\rho(-1/\tau)
= \sum_{\lam \in \Pplus}
\sum_{\rho \in \Exp}
{ \psi^*_{\a\pirho} S_{\lam\rho} \psi_{\b\pirho} \over S_{\id\rho} }
\chi_\lam(\tau)\,.
\ee
Hence, the coefficients of the open-string partition function
(\ref{eq:twopenpartition})
are given by
\be
\label{eq:opencoeff}
\nblama =  \sum_{\rho \in \Exp}
{ \psi^*_{\a\pirho} S_{\lam\rho} \psi_{\b\pirho} \over S_{\id\rho} }  \,.
\ee
Finally, the coefficients $\psi_{\a\pirho}$
relating the $\om$-twisted Cardy states
and $\om$-twisted Ishibashi states
may be identified
with the modular transformation matrices of characters
of twisted affine Lie algebras,
as may be seen, for example,
by examining the partition function of an open string
stretched between an $\om$-twisted and an
untwisted D-brane \cite{Ishikawa:2001zu}.

\section{Level-rank duality of twisted D-branes of $\suoddnk$ \label{A4} }
\renewcommand{\theequation}{D.\arabic{equation}}
\setcounter{equation}{0}

The finite Lie algebra $\suN$ possesses an order-two automorphism $\omc$
arising from the invariance of its Dynkin diagram under reflection.
This automorphism maps the Dynkin indices
of an irreducible representation $a_i \to a_{N-i}$,
and corresponds to charge conjugation of the representation.
This automorphism lifts to an automorphism of the affine Lie algebra $\suNK$,
leaving the \zeroth node of the extended Dynkin diagram invariant,
and gives rise to a class of $\omc$-twisted D-branes of
the $\suNK$ WZW model (for $N>2$).
Since the details of the $\omc$-twisted D-branes
differ significantly between even and odd $N$,
we will restrict our attention
to the $\omc$-twisted D-branes of the $\suoddnk = \Annonek$ WZW model.
An analysis of twisted D-branes for $\sunk$ for N and K $>$ 2  and untwisted  and twisted D-branes for $\sonk$ is in \cite{Naculich:2006mt}

First, recall that the $\omc$-twisted Ishibashi states $\Ctwishimu$
are labelled by self-conjugate integrable highest-weight
representations $\mu \in \Exp$ of $\Annonek$.
Equation (\ref{eq:integrable}) implies
that the Dynkin indices
$(a_0, a_1, a_2, \cdots, a_{n-1}, a_n, a_n, a_{n-1}, \cdots, a_1)$
of $\mu$ satisfy
\be
\label{eq:twistedintegrable}
a_0 + 2(a_1 + \cdots + a_n) = 2k+1 \,.
\ee
In ref.~\cite{Kacbuk}, it was shown that
the self-conjugate highest-weight representations of $\Annonek$
are in one-to-one correspondence with
integrable highest weight representations
of the associated orbit Lie algebra $\gcup = \Anntwok$,
whose Dynkin diagram is
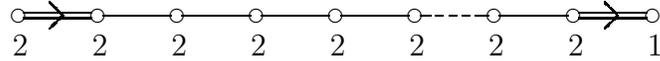
\begin{picture}(500,50)(10,10)
\put(100,40){\circle{5}}
\put(98,25){2}
\put(102,39){\line(1,0){26}}
\drawline(117,40)(112,45)
\drawline(117,40)(112,35)
\put(102,41){\line(1,0){26}}

\put(130,40){\circle{5}}
\put(128,25){2}
\put(132,40){\line(1,0){26}}

\put(160,40){\circle{5}}
\put(158,25){2}
\put(162,40){\line(1,0){26}}

\put(190,40){\circle{5}}
\put(188,25){2}
\put(192,40){\line(1,0){26}}

\put(220,40){\circle{5}}
\put(218,25){2}
\put(222,40){\line(1,0){26}}

\put(250,40){\circle{5}}
\put(248,25){2}
\put(252,40){\line(1,0){4}}
\put(258,40){\line(1,0){3}}
\put(263,40){\line(1,0){3}}
\put(268,40){\line(1,0){3}}
\put(273,40){\line(1,0){5}}

\put(280,40){\circle{5}}
\put(278,25){2}
\put(282,40){\line(1,0){26}}

\put(310,40){\circle{5}}
\put(308,25){2}
\put(312,41){\line(1,0){26}}
\drawline(327,40)(322,45)
\drawline(327,40)(322,35)
\put(312,39){\line(1,0){26}}

\put(340,40){\circle{5}}
\put(338,25){1}

\end{picture}

\option
with the integers indicating the dual Coxeter label $m_i$ of each node.
The representation $\mu \in \Exp$ corresponds
to the $\Anntwok$ representation $\pi(\mu)$ with Dynkin indices
$(a_0, a_1,  \cdots,  a_n)$.
Consistency with eq.~(\ref{eq:twistedintegrable})
requires that the dual Coxeter labels are
$(m_0, m_1, \cdots, m_n)=(1,2,2,\cdots,2)$,
and hence we must choose as the \zeroth node
the {\it right-most} node of the Dynkin diagram above.
The finite part of the orbit Lie algebra $\gcup$,
obtained by omitting the \zeroth node, is thus $C_n$.
(Note that $C_n$ is the orbit Lie algebra of the
finite Lie algebra $A_{2n}$ \cite{Kacbuk}.)

Observe that, by eq.~({\ref{eq:twistedintegrable}),
	$a_0$ must be odd,
	and that the representation $\pi(\mu)$ of the orbit algebra $\gcup$
	is in one-to-one
	correspondence \cite{Fuchs:1995zr,Kacbuk}
	with the integrable highest-weight representation $\pi(\mu)'$
	of the untwisted affine Lie algebra $\Conek$
	with Dynkin indices $(a_0', a_1', \cdots, a_n')$,
	where
	$a_0' = \half (a_0-1)$
	and $a_i' = a_i$ for $i=1, \cdots, n$.
	
	Next, the $\omc$-twisted Cardy states  $\Ccardya$
	(and therefore the $\omc$-twisted D-branes)
	of the $\Annonek$ WZW model
	are  labelled 
	by the integrable highest-weight representations
	$\a \in \Boundomc$
	of the twisted Lie algebra $\twhgk = \Anntwok$.
	We adopt these same conventions  for the labelling of the nodes
	of the Dynkin diagram.
	Thus, the Dynkin indices
	$(a_0, a_1, \cdots, a_n)$ of the highest weights $\a$
	must also satisfy eq.~(\ref{eq:twistedintegrable}),
	and the $\omc$-twisted D-branes are therefore characterized
	\cite{Kacbuk}
	by the irreducible representations of
	$C_n  = \spn$
	with Dynkin indices $(a_1, \cdots, a_n)$
	(also denoted, with a slight abuse of notation, by $\a$).
	That is, both the $\omc$-twisted Ishibashi states
	and the $\omc$-twisted Cardy states of $\suoddnk$
	are classified by integrable representations of $\spnk$.
	

	Recall from eq.~(\ref{eq:opencoeff}) that the coefficients
	of the partition function of open strings
	stretched between $\omc$-twisted D-branes $\a$ and $\b$
	are given by
	\be
	\nblama =  \sum_{\rho \in \Exp}
	{ \psi^*_{\a\pirho} S_{\lam\rho} \psi_{\b\pirho} \over S_{\id\rho} }
	\ee
	where $\a$, $\b \in \Boundomc$,  $\lam \in \Pplus$,
	and $\pirho$ is the representation
	of the orbit Lie algebra $\Anntwok$
	that corresponds to the self-conjugate representation $\rho$ of $\suoddnk$.
	The coefficients $\psi_{\a\pirho}$ are
	given \cite{Ishikawa:2001zu,Kacbuk}
	by the modular transformation matrix of the characters of $\Anntwok$.
	These in turn may be
	identified \cite{Ishikawa:2001zu,Fuchs:1995zr}
	with $\hS_{\a' \pirho'}$, the modular transformation matrix of
	$\Conek = \spnk$, so
	\be
	\label{eq:opencoefftwo}
	\nblama =  \sum_{\rho \in \Exp}
	{ \hS^*_{\a'\pirho'} S_{\lam\rho} \hS_{\b'\pirho'} \over S_{\id\rho} }  \,.
	\ee
	We will use this below to demonstrate level-rank duality of $\nblama$.

	\vs{.1in}
	\noindent{\bf Level-rank duality of the twisted open string spectrum}
	\vs{.1in}
	
	\option
	The coefficients of the partition function
	of open strings stretched between $\omc$-twisted D-branes  $\a$ and $\b$
	are real numbers so we may write (\ref{eq:opencoefftwo}) as
	\be
	\nblama =  \sum_{\rho \in \Exp}
	{ \hS_{\a'\pirho'} S^*_{\lam\rho} \hS^*_{\b'\pirho'}
		\over S^*_{\id\rho} }  \,.
	\ee
	Under level-rank duality,
	the $\suNK$ modular transformation matrices transform
	as \cite{Mlawer:1990uv}
	\be
	\label{eq:squrt}
	S_{\lam\mu}
	= \sqrt{K \over N} \e^{-2\pi i r(\lam) r(\mu)/NK} \tS^*_{\tlam\tmu}
	\ee
	and the (real) $\spnk$ modular transformation matrices transform
	as \cite{Mlawer:1990uv}
	\be
\label{eq:sss}
	\hS_{\a'\b'} = \htS_{\ta'\tb'} = \htSstar_{\ta'\tb'}
	\ee
	where $\tS$ and $\htS$ denote the $\suKN$ and $\spkn$
	modular transformation matrices respectively,
	$\tmu$ is the transpose of the Young tableau corresponding
	to the $\suNK$ representation $\mu$,
	and $\ta'$ is the transpose of the Young tableau corresponding
	to the $\spnk$ representation $\a'$.
	These imply
	\bea
	\label{eq:nblama}
	\nblama &=&  \sum_{\rho \in \Exp}
	{ \htSstar_{\ta'\widetilde{\pirho'} } \tS_{\tlam\trho}
		\htS_{\tb'\widetilde{\pirho'} } \over \tS_{\id\trho} }
	\e^{ 2\pi i r(\lam) r(\rho)/(2n+1)(2k+1)}
	\nonumber\\
	&=&
	\sum_{\rho \in \Exp}
	{ \tpsi^*_{\ta\widetilde{\pirho} } \tS_{\tlam\trho}
		\tpsi_{\tb\widetilde{\pirho} } \over \tS_{\id\trho} }
	\e^{ 2\pi i r(\lam) r(\rho)/(2n+1)(2k+1)}   \,.
	\eea
	Let $\rhohat$ be the self-conjugate $\suoddkn$ representation
	that maps to the $\spkn$ representation $\widetilde{\pirho'}$,
	which is the transpose of the $\spnk$ representation $\pirho'$.
	In other words, the representation $\pi(\rhohat)$ of the orbit algebra
	is identified with $\widetilde{\pirho}$.
	Now $\rhohat$ is not equal to $\trho$ (the transpose of $\rho$),
	which is generally not a self-conjugate representation,
	but they are in the same cominimal equivalence class,
	\be
	\label{eq:cominimal}
	\trho = \sig^{r(\rho)/(2n+1)} (\rhohat) .
	\ee
	Equation (\ref{eq:cominimal}) implies
	that \cite{Naculich:1990hg,Mlawer:1990uv}
	\be
	\tS_{\tlam\trho} =
	\e^{-2\pi i r(\lam) r(\rho)/(2n+1)(2k+1)}
	\tS_{\tlam\rhohat}
	\ee
	so that eq.~(\ref{eq:nblama}) becomes
	\be
	\nblama
	= \sum_{\rhohat}
	{ \tpsi^*_{\ta\pirhohat}  \tS_{\tlam\rhohat} \tpsi_{\tb\pirhohat}
		\over \tS_{\id\rhohat} }
	= \tn_{\tb\tlam}^{~~~\ta}   \,,
	\ee
	proving the level-rank duality of the coefficients of
	the open-string partition function of $\omc$-twisted D-branes
	of $\suoddnk$.
	That is,
	if the spectrum of an $\suoddnk$ open string
	stretched between $\omc$-twisted D-branes $\a$ and $\b$
	contains $\nblama$ copies of the
	highest-weight representation $V_\lam$ of $\suoddnk$,
	then the spectrum of an $\suoddkn$ open string
	stretched between $\omc$-twisted D-branes $\ta$ and $\tb$
	contains an equal number of copies of the
	highest-weight representation $V_{\tlam}$ of $\suoddkn$.

\newpage
\bibliographystyle{utphys}

\bibliography{emirefs}

\providecommand{\href}[2]{#2}\begingroup\raggedright
\endgroup

\end{document}